\documentclass[twocolumn,10pt]{tsfp}
\usepackage{flushend}
\usepackage{graphicx}
\usepackage[authoryear,round]{natbib}
\usepackage{fancyhdr}
\usepackage{amsmath}

\begingroup
\endgroup

\title{A formal $\log(Re)$-cost framework for the engineering turbulence problem}

\author{Jiaqi Li
    \affiliation{
        Mechanical Engineering \\
        The Pennsylvania State University \\
        University Park, PA 16802, USA \\
        jfl5663@psu.edu
    }
}

\author{Robert F. Kunz
    \affiliation{
        Mechanical Engineering \\
        The Pennsylvania State University \\
        University Park, PA 16802, USA\\
        rfk102@psu.edu
    }
}

\author{George Huang
    \affiliation{
        Mechanical and Materials Engineering \\
        Wright State University \\
        Dayton, OH 45435, USA \\
        george.huang@wright.edu
    }
}

\author{Xiang I.A. Yang
    \affiliation{
        Mechanical Engineering \\
        The Pennsylvania State University \\
        University Park, PA 16802, USA \\
        xzy48@psu.edu
    }
}

\begin{document}

\maketitle
\thispagestyle{fancy}
\fontsize{9}{11}\selectfont

\section*{ABSTRACT}

In fluid engineering, the turbulence problem is the longstanding challenge of obtaining accurate predictions of engineering quantities at affordable computational cost. Viewed through the lens of computational complexity, a practical algorithm requires cost growth no worse than $O(N)$, where $N$ denotes the problem size. For turbulent flows, the problem size may be approximated by the number of dynamically relevant scales and, therefore, by the Reynolds number $Re$. On this basis, the cost of a practical method should scale no worse than $O(Re)$. We propose a multi-fidelity, physics-constrained, data-driven framework designed to satisfy this criterion under the assumptions stated below. For demonstration, we augment the Spalart--Allmaras (SA) model through field inversion and machine learning (FIML) using a constrained formulation that preserves the law of the wall. The model is trained at a low Reynolds number, where high-fidelity data are affordable, and is then deployed at higher Reynolds numbers. Under the stated assumptions of a mean-flow-aligned grid for a wall-bounded flow, fixed spanwise resolution, and steady-solver cost linear in the grid-point count, the low-fidelity RANS prediction scales as $O(\log(Re))$. Relative to the target Reynolds number, the high-fidelity calculation and FIML stage each contribute $O(Re^0)$, so the overall formal cost scaling is $O(\log(Re))$. The framework is assessed in plane channel flow and in the periodic hill with slope factor $\alpha=1$. In channel flow, a model trained at $Re_\tau=1000$ corrects the wake-layer error of the baseline SA model and retains this improvement when extrapolated to $Re_\tau=5200$. In the periodic hill, a model trained at $Re_b=5600$ is tested at $Re_b=10595$, $19000$, and $37000$. The constrained formulation preserves the correct separation and recovery behavior as the Reynolds number increases. For comparison, we include results from conventional unconstrained FIML and from the baseline model. Across all tested Reynolds numbers, the constrained model yields the lowest root-mean-square error and exhibits nearly Reynolds-number-independent error, indicating robust extrapolation.

\section*{INTRODUCTION}

Turbulence remains one of the central unresolved problems in physics and engineering. In fluid engineering, the practical difficulty lies in predicting quantities such as drag and lift at high Reynolds numbers with both useful accuracy and acceptable computational cost. The core difficulty arises from the vast separation of dynamically relevant scales that must be represented.

Computational fluid dynamics (CFD) provides a hierarchy of modeling approaches, each reflecting a different compromise between accuracy and cost depending on which scales are resolved and which are modeled. Direct numerical simulation (DNS), which resolves all scales down to the Kolmogorov length, is regarded as the gold standard. However, even with optimized grids and algorithms, its cost scales as $Re_L^{2.91}$ \citep{yang2021gridb}, rendering it impractical beyond laboratory Reynolds numbers. Large-eddy simulation (LES) reduces this cost by modeling the smallest scales while resolving the inertial range, but still scales as $Re_L^{2.72}$. Wall-modeled LES (WMLES) and hybrid RANS/LES methods further reduce the cost to approximately $Re_L^{1.14}$. Despite their improved efficiency, these turbulence-resolving approaches remain too expensive for routine industrial applications, for example at $Re \sim 10^9$ \citep{slotnick2014cfd}. At the other end of the spectrum, Reynolds-averaged Navier--Stokes (RANS) models resolve only the mean flow while modeling all turbulent fluctuations and are therefore far more affordable than DNS and LES. Their weakness is accuracy: because there is no universal mapping between mean flow and turbulence, widely used closures such as Spalart--Allmaras (SA) and $k$--$\omega$ SST show limited performance in flows with strong separation and other non-equilibrium effects \citep{spalart2015philosophies,durbin2018some}.

To place this accuracy--cost trade-off in a broader context, it is useful to borrow the language of computational complexity. In computer science, algorithmic cost is assessed not only by the expense of solving a single instance but by how that expense grows with problem size $N$. Algorithms whose cost grows linearly or sublinearly remain practical as $N$ increases, whereas superlinear growth quickly becomes prohibitive in repeated use. A similar viewpoint may be adopted for turbulent flows. Because the Reynolds number measures the separation between the largest and smallest dynamically relevant scales, it may be interpreted as an effective measure of problem size. Cost scalings expressed in terms of $Re$ can therefore be viewed as statements of algorithmic complexity. From this perspective, the limitation of turbulence-resolving methods becomes immediate: their superlinear cost growth renders them impractical at engineering Reynolds numbers. RANS, by contrast, lies on fundamentally different computational footing because its cost depends only weakly on $Re$. A practical route forward is therefore to preserve the affordability of RANS while recovering some of the accuracy of high-fidelity methods.

The emergence of machine learning (ML) has opened new opportunities for turbulence modeling \citep{duraisamy2019turbulence}. Early developments primarily aimed to improve predictions for specific flow configurations \citep{wang2017physics,wu2018physics,parish2016paradigm}, whereas more recent efforts have focused on generalization across geometries and flow conditions \citep{rumsey2022search,srivastava2022generalizable,srivastava2024generalizably}. Among the available approaches, the field inversion and machine learning (FIML) framework is especially attractive because it augments an existing closure in a model-consistent manner \citep{singh2017machine,holland2019towards,rumsey2022search}. In the FIML framework, a baseline turbulence model is first recalibrated through field inversion so that it reproduces reference data, and a regression model is then trained to learn the relationship between local flow features and the inferred correction. Another closely related paradigm is progressive machine learning, which seeks to improve robustness and extrapolative capability without erasing successful calibrations already embedded in the baseline model \citep{bin2022progressive,bin2024two,bin2024constrained,yang2025data}. 

Despite these advances, most prior work in data-driven turbulence modeling has emphasized generalization in configuration space rather than extrapolation in Reynolds number. For engineering utility, however, Reynolds-number extrapolation is at least as important. Existing attempts at $Re$ extrapolation \citep{vadrot2023log} have shown promise in canonical flows such as channels and boundary layers, but their relevance to complex separated flows remains limited. This gap motivates the present study.

In this paper, we introduce a multi-fidelity, physics-constrained, data-driven turbulence-modeling framework developed specifically for extrapolation across Reynolds numbers. The framework combines three ingredients: the efficiency of RANS, whose cost depends only weakly on Reynolds number; the accuracy of high-fidelity data obtained at moderate $Re$; and the extrapolative capability of physics-constrained machine learning that preserves calibration with respect to the law of the wall. For demonstration, we augment the SA model using both constrained and unconstrained FIML formulations and assess the resulting models in plane channel flow and in the periodic hill \citep{breuer2009flow,rapp2011flow}. The goal is not universal cross-geometry generalization but reliable Reynolds-number extrapolation within a flow family while retaining the favorable cost structure of RANS.

\section*{RANS COST SCALING}

This section establishes the cost scaling for RANS simulations of boundary-layer flows. To the authors’ knowledge, such an explicit derivation is not commonly presented. Following \citet{yang2021gridb} and \citet{choi2012grid}, we consider a flow with a dominant streamwise direction $x$ and a wall-normal direction $y$, and assume that the grid is aligned with the mean-flow direction. This setting is representative of common RANS practice in wall-bounded flows.

Because RANS resolves only the mean flow, the grid spacing should be determined by the gradients of the mean velocity rather than by the smallest turbulent scales. A convenient estimate is
\begin{equation}
\Delta_x =\frac{c_xU_\infty}{\left|\partial U/\partial x\right|}\approx  \frac{c_x U_\infty}{|du_\tau/dx|},
\qquad
\Delta_y = \frac{c_y u_\tau}{\partial U/\partial y},
\label{eq:spacing}
\end{equation}
where $U_\infty$ is a characteristic outer velocity, $u_\tau$ is the friction velocity, and $c_x$ and $c_y$ are constants. The spanwise resolution is taken to be uniform, with a fixed number of points $N_z$, as is commonly done, for example, for bodies of revolution.

Under these assumptions, the total number of grid points is
\begin{equation}
N
= N_z \int\!\!\int \frac{dx\,dy}{\Delta_x \Delta_y}.
\label{eq:grid0}
\end{equation}
Substituting Eq.~\eqref{eq:spacing} into Eq.~\eqref{eq:grid0} gives
\begin{equation}
N
= \frac{N_z}{c_x c_y U_\infty}
\int\!\!\int
\frac{|du_\tau/dx|}{u_\tau}
\frac{\partial U}{\partial y}
\, dx\,dy.
\label{eq:grid1}
\end{equation}
Integrating first in the wall-normal direction yields
\begin{equation}
\int_0^{\delta} \frac{\partial U}{\partial y}\,dy
= U(\delta)-U(0)
\sim U_\infty,
\end{equation}
so that Eq.~\eqref{eq:grid1} reduces to
\begin{equation}
N
= \frac{N_z}{c_x c_y}
\int \frac{|du_\tau/dx|}{u_\tau}\,dx\sim \ln\left(\frac{u_{\tau,1}}{u_{\tau,2}}\right).
\label{eq:grid2}
\end{equation}

To express this result in terms of the Reynolds number, we use the standard relation
\begin{equation}
u_\tau = U_\infty \sqrt{\frac{C_f}{2}},
\end{equation}
together with the flat-plate turbulent boundary-layer estimate
\begin{equation}
C_f \sim Re_x^{-1/7}.
\end{equation}
It then follows that
\begin{equation}
N \sim \ln(Re).
\label{eq:logre}
\end{equation}

Equation~\eqref{eq:logre} shows that the RANS grid-point requirement grows only logarithmically with Reynolds number. If the cost of the steady solver scales linearly with the number of grid points, then the overall computational cost of RANS also scales as
\begin{equation}
\text{Cost}_{\mathrm{RANS}} \sim N\sim \log(Re).
\end{equation}

\section*{SA MODEL AND LAW OF THE WALL CONSTRAINT}
\label{sec:rbsa}

This section summarizes the baseline SA model and the law-of-the-wall (LoW)-constrained modification used in the present work, following \citet{spalart1992one} and \citet{bin2024constrained}. The transport equation for the modified eddy viscosity $\tilde{\nu}$ is
\begin{equation}
\begin{split}
\frac{\mathrm{D}\tilde{\nu}}{\mathrm{D}t}
= &\, c_{b1}\tilde{S}\tilde{\nu}
- c_{w1} f_w \left(\frac{\tilde{\nu}}{d}\right)^2 \\
&+ \frac{1}{\sigma}\left[
\frac{\partial}{\partial x_j}\left((\nu+\tilde{\nu})\frac{\partial \tilde{\nu}}{\partial x_j}\right)
+ c_{b2}\frac{\partial \tilde{\nu}}{\partial x_j}\frac{\partial \tilde{\nu}}{\partial x_j}
\right],
\end{split}
\label{eq:SA}
\end{equation}
where $\mathrm{D}/\mathrm{D}t$ denotes the material derivative. The eddy viscosity is given by $\nu_t=\tilde{\nu}f_{\nu1}$, and the anisotropic part of the Reynolds stress tensor is modeled with the Boussinesq hypothesis as $R_{ij}^d=2\nu_t S_{ij}$. The modified strain-rate magnitude is
\begin{equation}
\tilde{S}=\Omega+\frac{\tilde{\nu}}{\kappa^2 d^2}f_{\nu2},
\end{equation}
where $\Omega$ is the vorticity magnitude, $d$ is the distance to the nearest wall, and $\kappa=0.41$ is the von K\'arm\'an constant. The auxiliary functions are
\begin{equation}
f_{\nu1}=\left[1-\exp\left(-\frac{\chi}{\kappa A}\right)\right]^2,
~~~
f_{\nu2}=1-\frac{\chi}{1+\chi f_{\nu1}},
\end{equation}
where $\chi=\tilde{\nu}/\nu$, and the model coefficients are $A=17$, $c_{b2}=0.622$, and $\sigma=2/3$.
The present form of $f_{\nu1}$ differs from that in the original SA model \citep{spalart1992one}. It enforces the correct near-wall scaling, $f_{\nu1}\sim y^2$, while remaining consistent with the original SA formulation away from the wall. Finally, the function $f_w$ is defined as
\begin{equation}
f_w=g\left(\frac{1+2^6}{g^6+2^6}\right)^{1/6},
\qquad
g=r+0.3(r^6-r),
\end{equation}
where
$
r=\min\left({\tilde{\nu}}/({\tilde{S}\kappa^2 d^2}),\,10\right).
$

In the present study, the coefficient $c_{b1}$ is allowed to vary in space, following \citet{singh2017machine}, \citet{rumsey2022search}, and \citet{srivastava2024generalizably}. Two formulations are considered: a conventional unconstrained FIML approach and a constrained variant. In the unconstrained formulation, coefficients such as $c_{w1}$ remain fixed at their baseline values as $c_{b1}$ varies. In the constrained formulation, the variations of $c_{b1}$ and $c_{w1}$ are coupled so that the model remains consistent with the LoW. This preserves the correct near-wall asymptotic behavior during both field inversion and machine learning. Additional details are provided by \citet{bin2024constrained} and are not repeated here.

\section*{MULTI-FIDELITY FRAMEWORK}

The proposed framework consists of three stages. First, a high-fidelity dataset is generated at a low Reynolds number, where such calculations remain affordable. Second, FIML is used to augment the baseline turbulence model, which in the present study is the SA model. The model correction is introduced through a spatially varying $c_{b1}$. In the field-inversion step, an optimization problem is solved to infer the correction field that enables the augmented RANS solution to match the high-fidelity target data. In the subsequent machine-learning step, this inferred field is replaced by an explicit functional relation between local nondimensional flow features and the correction. Once trained, this relation can be evaluated directly at a new Reynolds number without further optimization, access to high-fidelity data, or case-by-case retuning. In the present implementation, field inversion is carried out with DAFoam \citep{he2020dafoam}, the regression model is a fully connected feedforward neural network, and the learned correction is deployed in OpenFOAM \citep{weller1998tensorial}. Third, the augmented closure is evaluated at higher Reynolds numbers using RANS.

Because the high-fidelity simulation and the learning stage are performed only once at a low Reynolds number, their cost does not scale with the target Reynolds number. The online prediction cost therefore inherits the favorable Reynolds-number scaling of RANS. In the language of algorithmic complexity, the overall cost is
\begin{equation}
\mathrm{Cost} = O(Re^0) + O(Re^0) + O(\log Re) = O(\log Re).
\end{equation}

Two formulations are considered: an unconstrained formulation and a constrained one. In the unconstrained formulation, all coefficients other than $c_{b1}$ retain their baseline values. In the constrained formulation, the coupling between $c_{b1}$ and the associated destruction term is modified so that the law of the wall is preserved. This distinction is important for Reynolds-number extrapolation. As the Reynolds number increases, the main structural change in many wall-bounded flows is the compression of the near-wall layer. A useful data-driven correction should therefore improve the outer-flow behavior without destroying the near-wall asymptotic calibration of the baseline model. The underlying assumption is that dominant nonequilibrium outer-layer physics, such as pressure-gradient and separation effects, can be learned at moderate Reynolds number and then transferred to higher Reynolds numbers, provided that the near-wall asymptotics are respected. This assumption is reasonable for many aerodynamic and hydrodynamic flows, in which increasing Reynolds number primarily thins the inner layer rather than introducing entirely new outer-layer physics.

Different feature sets are used for the two flow configurations considered here. For plane channel flow, the single feature $|P|/(|P|+|D|)$ is used, where $P$ and $D$ denote the production and destruction terms in the SA model. This feature isolates the wake-layer imbalance responsible for the baseline-model error. For the periodic hill, two features are used: $1/(f_w+1)$, which reflects the pressure-gradient effect, and $|P|/(|P|+|D|)$, which captures the balance between mean-shear production and near-wall destruction. The channel model is trained at $Re_\tau=1000$ using the DNS data of \citet{graham2016web}, and the periodic-hill model is trained at $Re_b=5600$.

\section*{RESULTS}

\subsection*{Plane channel flow}

\begin{figure*}[t]
\centering
\includegraphics[width=1\linewidth]{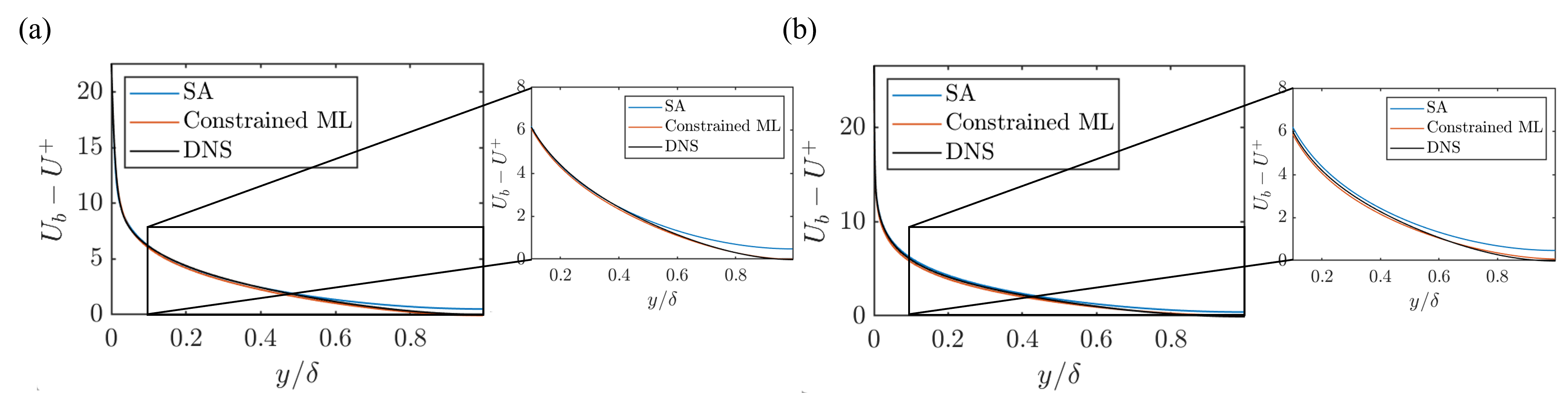}
\caption{Velocity-defect profiles in plane channel flow at (a) $Re_\tau=1000$ and (b) $Re_\tau=5200$.}
\label{fig:channel}
\end{figure*}

Plane channel flow provides a simple test of whether the framework can learn a correction at one Reynolds number and retain it at another. The baseline SA model reproduces the inner-layer behavior well, but it overpredicts the velocity deficit in the wake. Figure~\ref{fig:channel} shows that the constrained ML model reduces this wake-layer error at the training condition, $Re_\tau=1000$, and retains the improvement when extrapolated to $Re_\tau=5200$. This result is consistent with the expectation that the wake-layer correction should remain transferable when the near-wall scaling is preserved. The channel case therefore serves as a calibration check: if the framework cannot extrapolate in this comparatively simple flow, it is unlikely to succeed in a separated configuration.

\subsection*{Periodic hill}

The periodic hill provides a more demanding test because it involves separation, shear-layer development, and recovery, all of which vary with Reynolds number \citep{breuer2009flow,rapp2011flow}. Here we compare the baseline SA model, the unconstrained ML model, and the constrained ML model.

\begin{figure}[t]
\centering
\includegraphics[width=0.5\textwidth]{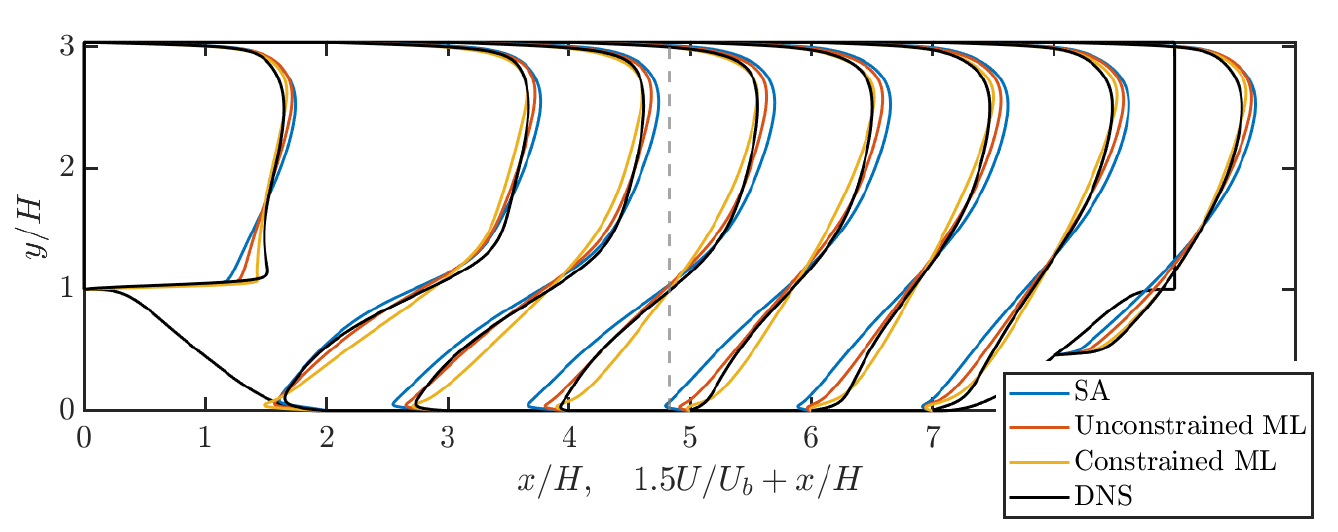}
\caption{Streamwise velocity profiles in the periodic hill at the training condition $Re_b=5600$. The dashed line marks the reattachment location.}
\label{fig:ph5600}
\end{figure}

\begin{figure}[t]
\centering
\includegraphics[width=0.45\textwidth]{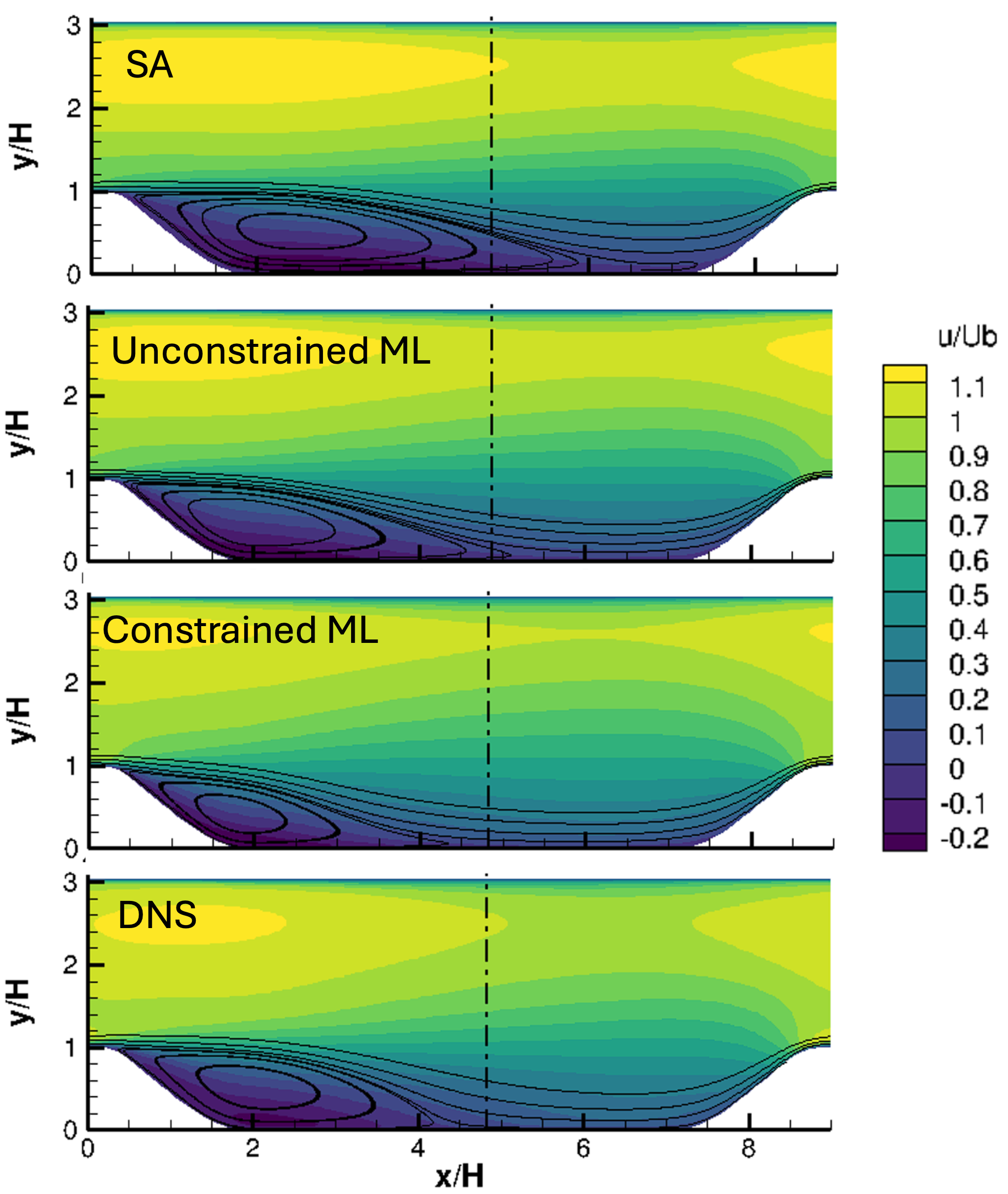}
\caption{Contours of streamwise velocity and streamlines at the training condition $Re_b=5600$. The dashed lines mark the reattachment locations.}
\label{fig:cont-ph5600}
\end{figure}

Figures~\ref{fig:ph5600} and \ref{fig:cont-ph5600} show the results at the training condition. Both machine-learned models improve the recirculation and recovery regions relative to the baseline SA model. The velocity profiles show better agreement with the reference data in and downstream of the separated region, and the contour plots indicate a more realistic recirculation bubble and recovery pattern. The difference between the constrained and unconstrained formulations remains modest at the training Reynolds number. This is expected because both models have sufficient flexibility to fit the low-Reynolds-number target case. At this stage, the training case shows primarily that both learned corrections can extract useful information from the reference data.

\begin{figure}[t]
\centering
\includegraphics[width=0.5\textwidth]{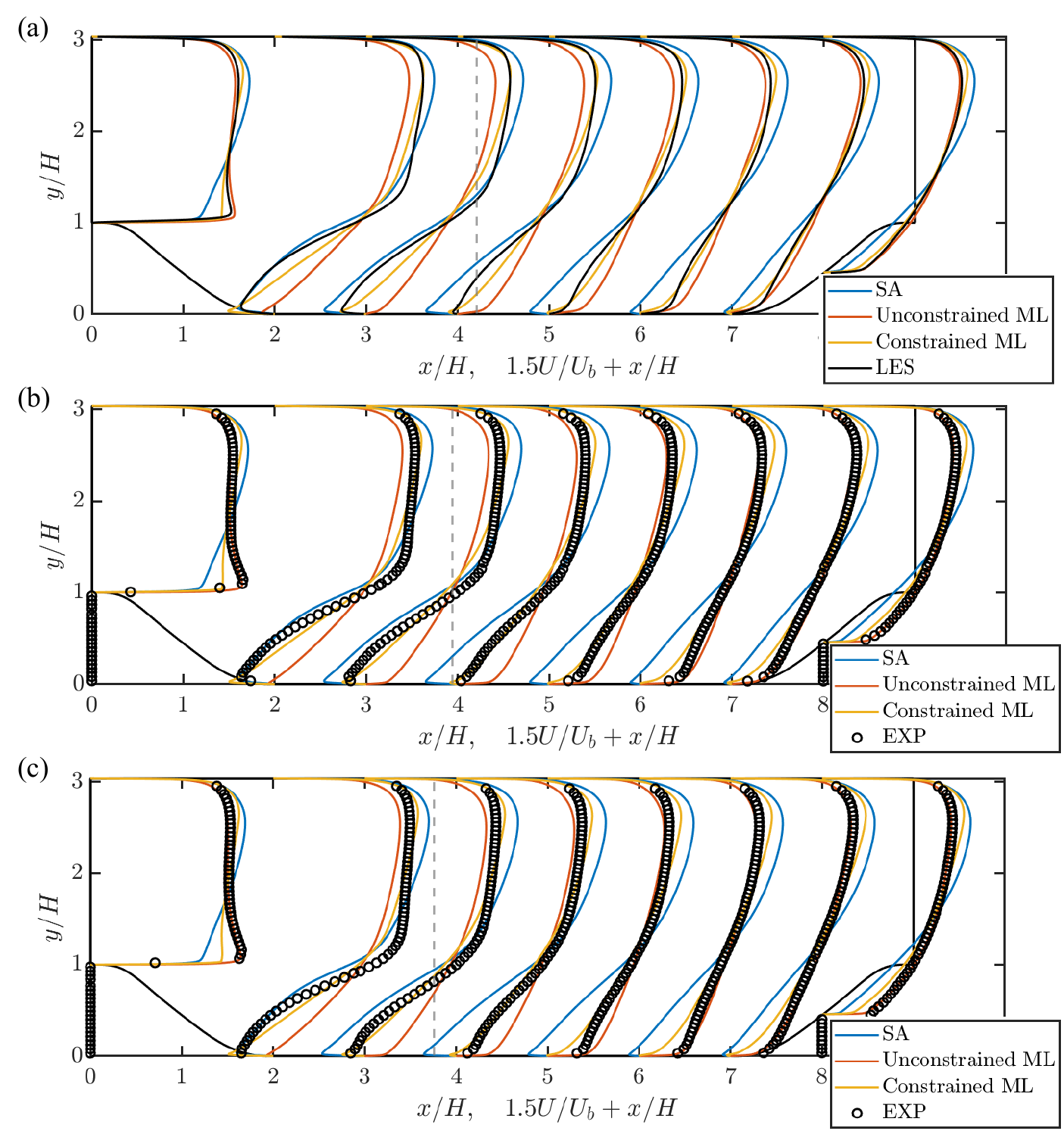}
\caption{Velocity profiles at test Reynolds numbers $Re_b=10595$, $19000$, and $37000$. Dashed lines mark the reattachment locations.}
\label{fig:phtest}
\end{figure}

\begin{figure*}[t]
\centering
\includegraphics[width=0.8\textwidth]{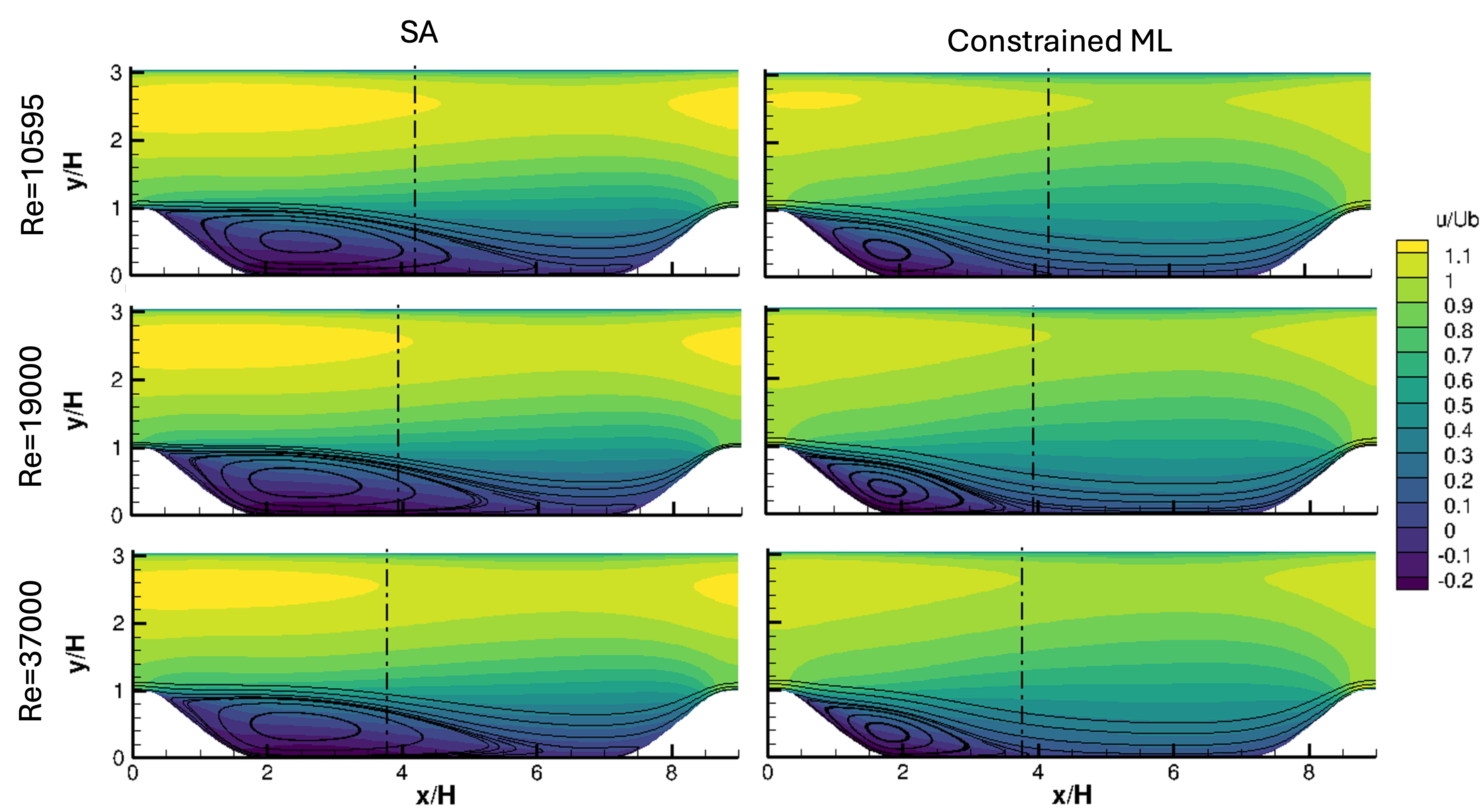}
\caption{Velocity contours and streamlines at test Reynolds numbers: (a) $Re_b=10595$, (b) $Re_b=19000$, and (c) $Re_b=37000$. Dashed lines mark the reattachment locations.}
\label{fig:cont_phtest}
\end{figure*}

The difference between the two ML formulations becomes much more pronounced during extrapolation. Figures~\ref{fig:phtest} and \ref{fig:cont_phtest} show the results at $Re_b=10595$, $19000$, and $37000$. The constrained ML model preserves the size of the separation bubble and predicts the downstream recovery much more accurately than either the baseline SA model or the unconstrained ML model. This behavior is visible in both the profile comparisons and the contour plots. The predicted reattachment location remains much closer to the reference value, and the overall mean-flow development downstream of reattachment is captured more faithfully.

In contrast, the unconstrained ML model predicts an unrealistically short separation region as the Reynolds number increases. Its performance at the training condition therefore does not translate into robust Reynolds-number extrapolation. This trend indicates that fitting the training case is not sufficient by itself; the augmentation must also preserve the near-wall structure required for extrapolation. The baseline SA model also deteriorates systematically with Reynolds number: the experimentally observed reattachment location shifts with $Re$, whereas the modeled reattachment location remains comparatively insensitive to $Re$.

\begin{figure}[t]
\centering
\includegraphics[width=0.9\linewidth]{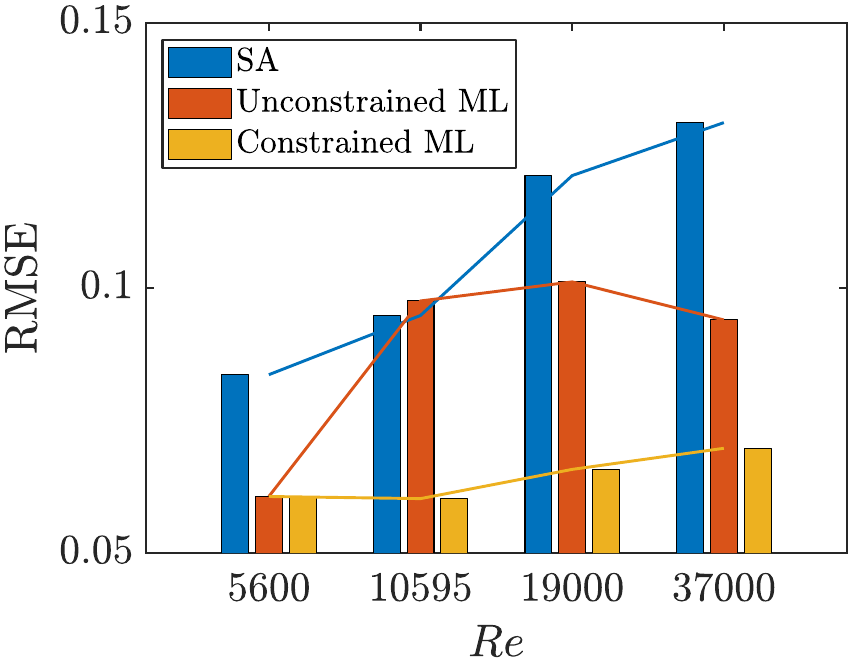}
\caption{RMSE of the periodic-hill velocity profiles relative to the experimental data.}
\label{fig:pherror}
\end{figure}

To quantify these trends, we compute the root-mean-square error (RMSE) of the streamwise velocity profiles relative to the experimental data:
\begin{equation}
\mathrm{RMSE} = \sqrt{\frac{1}{N}\sum_{i=1}^{N}\left(U_i-U_{i,\mathrm{ref}}\right)^2},
\label{eq:rmse}
\end{equation}
where the average is taken over the locations for which experimental data are available. The results are shown in Figure~\ref{fig:pherror}. The baseline SA model exhibits a monotonic increase in error with Reynolds number. The unconstrained ML model reduces the error relative to the baseline, but its error also grows with Reynolds number. By contrast, the constrained ML model yields the lowest RMSE for every test case and, importantly, maintains nearly constant error from $Re_b=5600$ to $Re_b=37000$. This behavior strongly suggests that the law-of-the-wall constraint is essential for stable Reynolds-number extrapolation in separated flows.

\section*{DISCUSSION AND CONCLUDING REMARKS}

The present results suggest that the engineering turbulence problem should be viewed as a balance between accuracy and computational complexity. The goal is not to eliminate modeling or to require scale-resolving simulation for every high-Reynolds-number prediction. Rather, the goal is to combine a computationally practical model with a calibration strategy that transfers the most useful high-fidelity information to the regime of interest. In the present framework, the expensive steps---high-fidelity simulation and field inversion---are performed at a low or moderate Reynolds number, while predictions at the target high Reynolds number retain the cost structure of RANS.

This perspective also clarifies why Reynolds-number extrapolation differs from cross-geometry generalization. Cross-geometry generalization requires the model to accommodate different dominant mechanisms and flow organizations. Reynolds-number extrapolation, by contrast, often preserves the large-scale structure of the flow while changing the thickness of the inner layer and the degree of scale separation. For this reason, preserving the law of the wall is especially important. The periodic-hill results support this view: both learned models improve the training case, but only the constrained model remains robust as the Reynolds number increases.

The framework also suggests a practical workflow for engineering applications. One may train on a canonical or reduced configuration for which high-fidelity data are feasible, use field inversion to expose the dominant model discrepancy, and then learn a constrained correction for deployment over a family of Reynolds numbers. This strategy is not universally applicable, because increasing Reynolds number may introduce new physics not represented in the training case. Nevertheless, for many attached and separated vehicle-relevant flows, this requirement is more realistic than demanding universal cross-geometry extrapolation from a single dataset.

The present results should therefore be viewed as an existence proof rather than a universal closure. For both a canonical attached flow and a separated flow, the proposed framework retains RANS-level online cost, preserves the correct wall scaling, and substantially improves extrapolative behavior through data-enabled augmentation.

In summary, this work examined Reynolds-number extrapolation in data-driven turbulence modeling using a physics-constrained FIML framework. The learned correction reduces the wake-layer bias of the SA model in plane channel flow and remains effective when applied at the higher Reynolds number. In the periodic hill, both the constrained and unconstrained ML models improve predictions for the training case, but only the constrained formulation remains reliable as the Reynolds number increases. Across all periodic-hill cases, the constrained model yields the lowest RMSE and nearly Reynolds-number-independent error. These results show that, for Reynolds-number extrapolation, model consistency matters as much as data fitting. More broadly, they suggest that computational complexity, physics-based closure design, and machine learning must be considered together if a practical $\log(Re)$ solution is to be achieved.

\section*{ACKNOWLEDGMENTS}

Yang acknowledges AFOSR Grant No. FA9550-23-1-0272 with Dr. Gregg Abate as the technical monitor.

\bibliographystyle{tsfp}
\bibliography{sample}

@string{jcp = {J. Comput. Phys.}}

@string{aiaa = {AIAA J.}}

@string{arfm = {Annu. Rev. Fluid Mech.}}

@string{jfm = {J. Fluid Mech.}}

@string{pof = {Phys. Fluids}}

@string{prf = {Phys. Rev. Fluids}}

@string{pas = {Prog. Aerosp. Sci.}}

@string{prf = {Phys. Rev. Fluid}}

@string{taml = {Theor. Appl. Mech. Lett.}}

@article{durbin2018some,
  title={Some recent developments in turbulence closure modeling},
  author={Durbin, Paul A},
  journal=arfm,
  volume={50},
  pages={77--103},
  year={2018},
  publisher={Annual Reviews},
  doi = {10.1146/annurev-fluid-122316-045020}
}

@inproceedings{rumsey2022search,
  title={In search of data-driven improvements to {RANS} models applied to separated flows},
  author={Rumsey, Christopher L and Coleman, Gary N and Wang, Li},
  booktitle={AIAA Scitech 2022 Forum},
  pages={0937},
  year={2022},
  doi={10.2514/6.2022-0937}
}

@article{wu2018physics,
  title={Physics-informed machine learning approach for augmenting turbulence models: A comprehensive framework},
  author={Wu, JinLong and Xiao, Heng and Paterson, Eric},
  journal=prf,
  volume={3},
  number={7},
  pages={074602},
  year={2018},
  publisher={APS},
  doi={10.1103/PhysRevFluids.3.074602}
}

@article{wang2017physics,
  title={Physics-informed machine learning approach for reconstructing {Reynolds} stress modeling discrepancies based on DNS data},
  author={Wang, Jianxun and Wu, Jinlong and Xiao, Heng},
  journal=prf,
  volume={2},
  number={3},
  pages={034603},
  year={2017},
  publisher={APS},
  doi={10.1103/PhysRevFluids.2.034603}
}

@article{spalart2015philosophies,
  title={Philosophies and fallacies in turbulence modeling},
  author={Spalart, Philippe R},
  journal=pas,
  volume={74},
  pages={1--15},
  year={2015},
  publisher={Elsevier},
  doi={10.1016/j.paerosci.2014.12.004}
}

@inproceedings{spalart1992one,
  title={A one-equation turbulence model for aerodynamic flows},
  author={Spalart, Philippe and Allmaras, Steven},
  booktitle={30th Aerospace Sciences Meeting and Exhibit},
  pages={439},
  year={1992},
  doi={10.2514/6.1992-439}
}

@article{bin2022progressive,
  title={Progressive, extrapolative machine learning for near-wall turbulence modeling},
  author={Bin, Yuanwei and Chen, Lihua and Huang, George and Yang, Xiang I A},
  journal=prf,
  volume={7},
  number={8},
  pages={084610},
  year={2022},
  publisher={APS},
  doi={10.1103/PhysRevFluids.7.084610}
}

@article{choi2012grid,
  title={Grid-point requirements for large eddy simulation: {Chapman’s} estimates revisited},
  author={Choi, Haecheon and Moin, Parviz},
  journal=pof,
  volume={24},
  number={1},
  year={2012},
  publisher={AIP Publishing}
}

@article{yang2021gridb,
  title={Grid-point and time-step requirements for direct numerical simulation and large-eddy simulation},
  author={Yang, Xiang I. A. and Griffin, Kevin P},
  journal={Physics of Fluids},
  volume={33},
  number={1},
  year={2021},
  publisher={AIP Publishing}
}

@article{bin2024constrained,
author = {Bin, Yuanwei and Huang, George and Kunz, Robert and Yang, Xiang I. A.},
title = {Constrained Recalibration of {Reynolds-averaged Navier–Stokes} Models},
journal = aiaa,
volume = {62},
number = {4},
pages = {1434-1446},
year = {2024},
doi = {10.2514/1.J063407},
}

@article{parish2016paradigm,
  title={A paradigm for data-driven predictive modeling using field inversion and machine learning},
  author={Parish, Eric J and Duraisamy, Karthik},
  journal=jcp,
  volume={305},
  pages={758--774},
  year={2016},
  publisher={Elsevier},
  doi={10.1016/j.jcp.2015.11.012}
}

@article{singh2017machine,
  title={Machine-learning-augmented predictive modeling of turbulent separated flows over airfoils},
  author={Singh, Anand Pratap and Medida, Shivaji and Duraisamy, Karthik},
  journal=aiaa,
  volume={55},
  number={7},
  pages={2215--2227},
  year={2017},
  publisher={American Institute of Aeronautics and Astronautics},
  doi={10.2514/1.J055595}
}

@inproceedings{holland2019towards,
  title={Towards integrated field inversion and machine learning with embedded neural networks for {RANS} modeling},
  author={Holland, Jonathan R and Baeder, James D and Duraisamy, Karthik},
  booktitle={AIAA Scitech forum},
  pages={1884},
  year={2019},
  doi={10.2514/6.2019-1884}
}

@article{weller1998tensorial,
  title={A tensorial approach to computational continuum mechanics using object-oriented techniques},
  author={Weller, Henry G and Tabor, Gavin and Jasak, Hrvoje and Fureby, Christer},
  journal={Comput. phys.},
  volume={12},
  number={6},
  pages={620--631},
  year={1998},
  publisher={American Institute of Physics},
  doi={10.1063/1.168744}
}

@phdthesis{srivastava2022generalizable,
  title={Generalizable Data-driven Model Augmentations Using Learning and Inference assisted by Feature-space Engineering},
  author={Srivastava, Vishal},
  year={2022},
  school={University of Michigan},
  url={https://dx.doi.org/10.7302/6032}
}

@article{duraisamy2019turbulence,
  title={Turbulence modeling in the age of data},
  author={Duraisamy, Karthik and Iaccarino, Gianluca and Xiao, Heng},
  journal=arfm,
  volume={51},
  pages={357--377},
  year={2019},
  publisher={Annual Reviews}
}

@inproceedings{srivastava2024generalizably,
  title={On Generalizably Improving {RANS} Predictions of Flow Separation and Reattachment},
  author={Srivastava, Vishal and Rumsey, Christopher L and Coleman, Gary N and Wang, Li},
  booktitle={AIAA SCITECH 2024 Forum},
  pages={2520},
  year={2024}
}

@techreport{slotnick2014cfd,
  title={{CFD} vision 2030 study: a path to revolutionary computational aerosciences},
  author={Slotnick, Jeffrey P and Khodadoust, Abdollah and Alonso, Juan and Darmofal, David and Gropp, William and Lurie, Elizabeth and Mavriplis, Dimitri J},
  year={2014},
  institution={NASA}
}

@article{bin2024two,
  title={Constrained re-calibration of two-equation {Reynolds-averaged Navier--Stokes} models},
  author={Bin, Yuanwei and Hu, Xiaohan and Li, Jiaqi and Grauer, Samuel J and Yang, Xiang I. A.},
  journal=taml,
  volume={14},
  number={2},
  pages={100503},
  year={2024},
  publisher={Elsevier}
}

@article{graham2016web,
  title={A web services accessible database of turbulent channel flow and its use for testing a new integral wall model for LES},
  author={Graham, J and Kanov, K and Yang, X I. A. and Lee, M and Malaya, N and Lalescu, CC and Burns, R and Eyink, G and Szalay, A and Moser, RD and others},
  journal={J. Turbul.},
  volume={17},
  number={2},
  pages={181--215},
  year={2016},
  publisher={Taylor \& Francis}
}

@article{yang2025data,
  title={Data-enabled discovery of specific and generalisable turbulence closures},
  author={Yang, Zhongxin and Shan, Xianglin and Yang, Xiang I. A. and Zhang, Weiwei},
  journal=jfm,
  volume={1016},
  pages={R1},
  year={2025},
  publisher={Cambridge University Press}
}

@article{vadrot2023log,
  title={Log-law recovery through reinforcement-learning wall model for large eddy simulation},
  author={Vadrot, Aur{\'e}lien and Yang, Xiang I. A. and Bae, H Jane and Abkar, Mahdi},
  journal=pof,
  volume={35},
  number={5},
  year={2023},
  publisher={AIP Publishing}
}

@article{he2020dafoam,
  title={Dafoam: An open-source adjoint framework for multidisciplinary design optimization with openfoam},
  author={He, Ping and Mader, Charles A and Martins, Joaquim RRA and Maki, Kevin J},
  journal={AIAA Journal},
  volume={58},
  number={3},
  pages={1304--1319},
  year={2020},
  publisher={American Institute of Aeronautics and Astronautics}
}

@article{rapp2011flow,
  title={Flow over periodic hills: an experimental study},
  author={Rapp, Ch and Manhart, M},
  journal={Experiments in fluids},
  volume={51},
  number={1},
  pages={247--269},
  year={2011},
  publisher={Springer}
}

@article{breuer2009flow,
  title={Flow over periodic hills--numerical and experimental study in a wide range of {R}eynolds numbers},
  author={Breuer, Michael and Peller, Nikolaus and Rapp, Ch and Manhart, Michael},
  journal={Computers \& Fluids},
  volume={38},
  number={2},
  pages={433--457},
  year={2009},
  publisher={Elsevier}
}

\end{document}